\documentclass[onecolumn, prc,nofootinbib]{revtex4-2}
\usepackage{epsfig,graphicx,float,pst-all}
\usepackage{amssymb}
\usepackage{hyperref}
\usepackage{physics}

\newcommand{\be}{\begin{equation}} 
\newcommand{\ee}{\end{equation}}
\newcommand{\bc}{\begin{center}}
\newcommand{\ec}{\end{center}}

\begin{document}
\title{Jetter and Post nuclear fusion cycles: new fire to an old idea}
\author{Lorenzo Fortunato$^{1,2}$}
\author{Andres Felipe Lopez Loaiza$^1$}
\author{Giulio Albertin$^1$}
\author{Enrico Fragiacomo$^3$}
\affiliation{1) Dip. Fisica e Astronomia ``G.Galilei", Univ. Padova (Italy)\\
	2) I.N.F.N. Sez. Padova, Padova (Italy)\\
	3) I.N.F.N. Sez. Trieste, Trieste (Italy)}

\begin{abstract}
New calculations of the time evolution and isotopic composition for a network of nuclear reactions breathe new life into an old idea in nuclear fusion, burning solid room temperature lithium-6 deuteride ($^6$LiD) with neutrons. Modern-day compilations of nuclear cross-sections are nowadays available and we use them to predict the full course of networks of thermonuclear reactions, re-examining the Jetter (n+$^6$Li) and Post cycles (p+$^6$Li), that offer great prospects for energy production in devices not based on plasma confinement. We present ideal
calculations, i.e. not including the energy loss due to the Stopping power, and more realistic calculations that include the Bethe-Bloch formula. We find that a significant amount of energy can, in principle, be generated.
\end{abstract}

\maketitle

\section{Fusion cycles and some history}
With modern techniques, we re-examine an almost forgotten concept that has gone lost amid unpublished Cold War era documents, finding a great potential for new fusion devices that are not based on creating a plasma of any sort.

In the '50's, in connection with warfare uses of nuclear energy, U. Jetter \cite{Jett} considered the following cycle of binary nuclear reactions:
\be
\begin{tabular}{lcccccccr}
	&$n$ & $+$ & \underline{$^{6}\mathrm{Li}$} & $\rightarrow$ & $\alpha$ & $+$ & $\mathrm{T}$ & +4.8 MeV\\
	&$\uparrow$ &  &  &  &  &  & $\downarrow$  & \\
	17.6 MeV +&$n$ & $+$ & $\alpha$ & $\leftarrow$ & \underline{$\mathrm{D}$} & $+$ & $\mathrm{T}$& 
\end{tabular}
\label{Jet}
\ee
as a way of liberating energy from fusion chain reactions, provided an external source of neutrons, that act as catalysts. In the equations, the target particles (not moving) have been underlined. R.F. Post (Livermore) \cite{Post}, considered instead the mirror cycle initiated by protons (also mentioned in Ref. \cite{McNal4})
\be
\begin{tabular}{lcccccccr}
	& $p$ & $+$ & \underline{$^{6}\mathrm{Li}$} & $\rightarrow$ & $\alpha$ & $+$ & $^{3}\mathrm{He}$ & +4.0 MeV \\
	&$\uparrow$ &  &  &  &  &  & $\downarrow$ & \\
	18.4 MeV +&$p$ & $+$ & $\alpha$ & $\leftarrow$ & \underline{$\mathrm{D}$} & $+$ & $^{3}\mathrm{He}$ & 
\end{tabular}
\label{Pos}
\ee
which is slightly disfavoured with respect to the previous cycle, due to the Coulomb barrier.
Either of the two primary reactions release quick suprathermal $^3$H or $^3$He ions that, upon a secondary reaction with $D$, return the initial catalyst with a different final energy. These final neutrons or protons are ready to start new reactions with the specific cross-sections at their respective new energies. Unlike fission, where neutrons are multiplied at each generation, the Jetter cycle is a fusion chain reactions without proliferation of neutrons, giving an ideally unitary neutron reproduction factor $k=1$. Inefficiencies, loss at the boundaries and other processes move the balance below criticality, $k<1$.

These reactions are crucially interesting because of the following properties: i) sub-criticality (neutron multiplication is not present); ii) self-termination within fractions of a second whenever the external driving neutron flux is stopped; iii) intrinsical radiological safety, as the main outcome are $^4$He and traces of light nuclei (except for possible loss of scattered neutrons from the reaction chamber); iv) no dependence on tritium economy, as this fuel material is obtained from lithium.

The Jetter cycle is sometimes mentioned in conjunction with thermonuclear weapon studies \cite{Makh} or with tritium breeding in the blanket of a fusion reactor \cite{Braa,Rubel}, where escaping neutrons are converted into fuel material.
Other studies have evidenced the special role of this isotope in various contexts: for instance Ref.\cite{Grig} analyzes the main reactions involving $^6$Li and compares Accelerator Driven Systems (ADS) with traditional fission reactors. $^6$Li has also been proposed as a dopant to enhance the yield of D+D fusion reactions in a plasma via fast neutron production \cite{Chir}. 

The discussion about the feasibility of using these cycles for energetic purposes started quite early, but it was kept secret or unpublished for a long time and then resumed mainly by J. Rand McNally Jr. in the 70’s \cite{McNal1,McNal2,McNal3,McNal4,McNal5,McNal6,McNal7}. He proposed that suprathermal ions in plasmas confined in magnetic mirror devices could start the cycles (\ref{Jet}), (\ref{Pos}) and other cycles and speculated if they could be effectively used for extraction of nuclear energy.
Throughout the discussions, the author expressed several times the idea that the
poor data set known for cross sections at that time had hindered the possible decisions
of choosing a specific fuel and the predictions of the energy generated that might
be converted into electricity. This is still partly true today, especially for reactions involving tritium.

\begin{figure}[t!]
	\begin{center}
		\includegraphics[clip=,width=.5\textwidth]{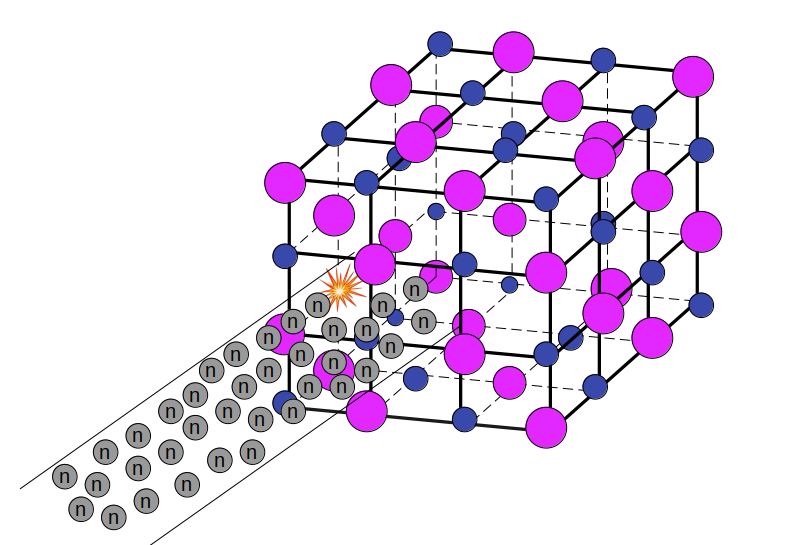}
	\end{center}
	\caption{In our scheme, a neutron beam impinges on a $^6$LiD crystal, thereby initiating a nuclear fusion reaction cycle.}
	\label{cryst}
\end{figure}

We re-examine here the reaction cycles (\ref{Jet}) and (\ref{Pos}), by using state-of-the-art cross-section measurements, we calculate reaction yields and abundance evolution for various set-ups and show that these processes might be ideal candidates for accelerator driven nuclear fusion reactors utilizing beams of neutrons or, for the rapidly growing field of laser-driven neutron sources (LDNS). In most of the discussions by Rand McNally and colleagues, they had in mind hot plasmas. In our scheme instead, cold solid (crystal or powder) $^6$LiD in a reactor is exposed to neutron radiation (see depiction in Fig. \ref{cryst}). The ``nuclear fire" \cite{McNal4,McNal6}, in which released particles from the irradiated zone reach out new surrounding fuel is analogous to a controlled chemical combustion.
While the relevance of $^6$Li has been theoretically underlined, a full incorporation of reaction dynamics was recognized and demanded already about half a century ago \cite{McNal1}.
It is imperative to know the evolution of a number of concurrent nuclear reactions in order to estimate the total yields, the burning time, the produced power and the fate of intermediate or final products.
The old references cited so far lacked a comprehensive computational approach to the solution of a net of reactions and got past of it with a simplified approach with a single equation for the time evolution of the catalyst (see for instance \cite{McNal5}). Our aim is to fill this gap by solving the set of coupled differential equations for the isotopic abundancies as a function of time.
In this sense, we do not need the machinery of plasma physics such as for instance Fokker-Planck kinetic simulations, transport equations and the like.
We rather study the fusion process {\it in solido} where different initial conditions might affect these exoenergetic reactions (and eventually lead to the destruction of the fuel) and find that they are well-suited to give appreciable thermal energy that should be further converted with traditional methods. For this scheme, the generation of a plasma is an accidental transient event, rather than a necessary condition.
The closest conceptualization that we have found in literature is that of ``wet-wood burners" of Dawson, Furth and Tenney \cite{Daws, McNal4}. The burning of this ``nuclear controlled fire" might happen in a small irradiated portion of a larger fuel material and a relatively small magnetic field might keep most of the slowest charged particles confined in the bulk.
A. J. Fisher patented in West Germany and in the U.S.A. \cite{Fish} the concept for a device, sometimes called FISCATRON, whereby electrical discharge initiated microexplosions should burn a solid fuel material of tritiated 6-lithium deuteride, $^6$LiD$_{1-x}$T$_x$, or 6-lithium deuterotritide. The patent proposes to use intense pulsated currents to explode a conducting wire surrounded by fusion fuel materials, although it is not presently clear if this approach can be made to work. In addition, the patent did not specify what is the expected evolution of the nuclear reactions, which is exactly the focus of this paper. 

Surprisingly, despite the enormous importance of this topic for the possible implications in energy production and for the economical and societal issues involved in it, a detailed analysis of these thermonuclear reactions is not available. We presently try to address this problem. After introducing terminology, inputs and methods, we present results for the ideal case in Sect. \ref{ideal} with the aim of completing the 50 years old challenge to a detailed description of reaction dynamics, and we show in Sect. \ref{real} realistic calculations that include the stopping effect  of charged particles in the bulk of fuel. Clearly, the ideal case shows a huge energy release that is then considerably resized in the realistic case. Further studies should be based on the latter case.

\subsection{Other reactions}
We have limited our initial speculations to the main reactions of the two cycles (Jetter's  and Post's), but about a hundred other side reactions are possible \cite{McNal5}, mostly exothermic and some of them with appreciable rates at the energies of interest. One that is particularly relevant,  $^3$He+$^6$Li $\rightarrow$ p+2$\alpha$ +16.88 MeV has been analyzed by McNally in Ref. \cite{McNal6} in connection with the $^6$Li+p $\rightarrow$ $^3$He + $\alpha$ +4.02 MeV reaction, as a way to burn lithium with protons to $3\alpha$ particles. This gives room for further modern explorations of this fascinating topic.

\section{Inputs}
\begin{figure}[t!]
 	\begin{center}
    \includegraphics[clip=,width=.98\textwidth]{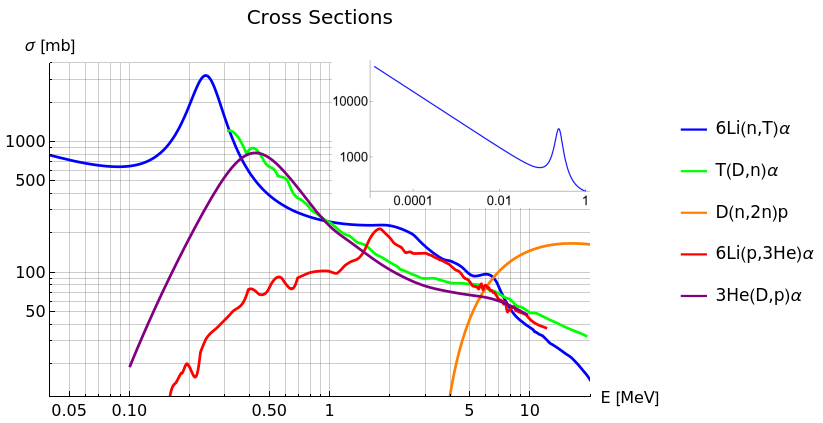}
 	\end{center}
 	\caption{Cross-sections for the reaction used in the code as a function of projectile energy in bilogarithmic scale. Interpolation smoothness varies depending on datasets. Data from \protect\cite{exfor,endf}. Notice that the blue curve cross-section grows linearly at low neutron energy (see insert).}
 	\label{cs}
\end{figure}

We take into account all the reactions in Eqs. (\ref{Jet}) and (\ref{Pos}) plus a reaction that couples the two cycles, i.e. the endothermic neutron-induced deuteron breakup reaction:
\be
\underline{\mathrm{D}}+n \rightarrow 2n+p - 2.2245~ \mathrm{MeV}
\label{knock}
\ee
This reaction triggers the Post cycles, when the energetic neutrons hit deuterium rather than lithium. These neutrons might be part the beam (if it is made energetic enough to overcome the threshold) or from the DT reaction in (\ref{Jet}).
The experimental inputs to our calculations are state-of-the art cross-section data compilations and interpolations as a function of the center of mass energy \cite{exfor, endf} plus atomic masses and binding energies data \cite{iaea}.
The cross-sections for the processes of interest have been interpolated and, when necessary because of the presence of multiple datasets at the same energy, a weighted average has been taken. They are collected in a single plot in Fig. \ref{cs}.
 
\section{Methodology} 
We have built a code for Mathematica \cite{Code} that takes as an input cross-section data for the network of reactions listed above and calculate the evolution of abundance of each species $Y_i$ with time, by solving a set of coupled first order nonlinear differential equations of the form:
\begin{equation}
dY_{i}/dt = +Y_{j}Y_{k}(\rho N_A \ev{\sigma v}_{jk,i})-Y_{i}Y_{j}(\rho N_A \ev{\sigma v}_{ij,n})+Y_{l}\lambda_{l}-Y_{i}\lambda_{i}+ \ldots
\label{set}
\end{equation}
where $\rho$ is the total density, $\lambda$ are decay constants, $N_A$ is the Avogadron number and $\ev{\sigma v}$ are reactions rates.
Notice that a species might be produced by these processes ($+$ sign) or depleted ($-$).
Also note that, while this is a common approach, for example in astrophysics or in confined plasma, where one might assume thermal velocity distributions and thus average over Maxwellian functions, this is not true here. Beam particles are injected with a constant speed and rate and fragments of binary reactions do share the Q-value energy according to their masses, emerging always with a certain given kinetic energy. They will remain suprathermal for a relatively long time, thus we can assume Kronecker $\delta$ velocity distributions at the formation.
They will most probably undergo a second collision before having degraded appreciably their kinetic energy via ionization, etc. The effect of losses is neglected in Sect. \ref{ideal} and then implemented and discussed in Sect. \ref{real}.

Different is the case for the reaction (\ref{knock}), where the three final fragments with almost the same mass, share the released energy. We will assume that the incoming neutron inelastically scatters, loosing an energy that breaks the deuteron binding with uniform energy-distribution. There are more refined ways to take into account the breakup into continuum states of deuterium, but this is not going to affect our calculations too much.
Very importantly, the first equation (see supplemental material) for the neutrons of the beam has a source term, $i$, that represents the incoming injection rate. This can be a constant or made to vary in time (for example bunched injections might be modeled).

\section{Results for the ideal case}\label{ideal}
\begin{figure}[t!]
	\begin{center}
	\includegraphics[clip=,width=.7\textwidth]{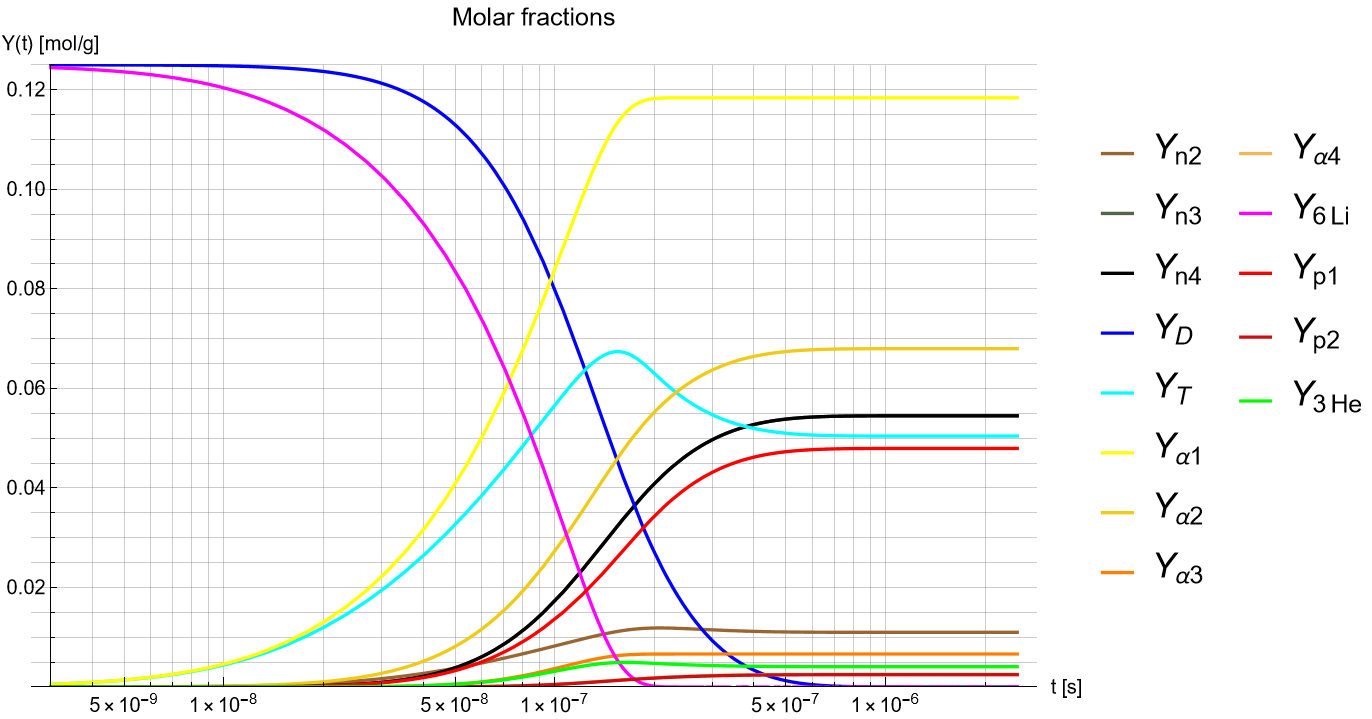}
	\end{center}
	\caption{Evolution of molar fraction (linear vertical scale) of each species with time for a bulk of cold $^6$LiD subject to a flux of 10$^{6}$ neutrons/sec with energy $E_{n_1}= 0.24$ MeV. }
	\label{evo}
\end{figure}

The typical outcome for a given injection energy of $E_{n_1}= 0.24$ MeV (i.e. relative maximum of the blue curve in Fig. \ref{cs}) and neutron flux, $10^6$ neutron/s is shown Fig. \ref{evo}, where one can follow the evolution in time of the molar fractions (or better nucleon fractions).
The dynamics is exhausted in less than a microsecond, leaving mostly $\alpha$ particles, some tritium plus some neutrons and  protons. Clearly, this is an upper limit, in the sense that we are considering all particles as recirculating through the same bulk material (equivalent to periodic boundary conditions). In reality, losses at the surfaces occur, but we might forget about this problem, considering a bulk of fuel material submersed in other fuel, even though the latter is not subjected to the action of the neutron beam.
Our results depend of course on two important ingredients, the intensity of the impinging neutron flux (affecting timescales) and the kinetic energy of the neutrons (affecting cross-sections and rates of the initial processes). Similar plots can be produced by varying these two variables. 
A relevant parameter to assess the efficiency of the burning process is the relative conversion into $\alpha$ particles at the end of the process, i.e. $\sum Y_{\alpha_i} (t_{final})$. 

\begin{figure}[t!]
	\begin{center}
	\includegraphics[clip=,width=.7\textwidth]{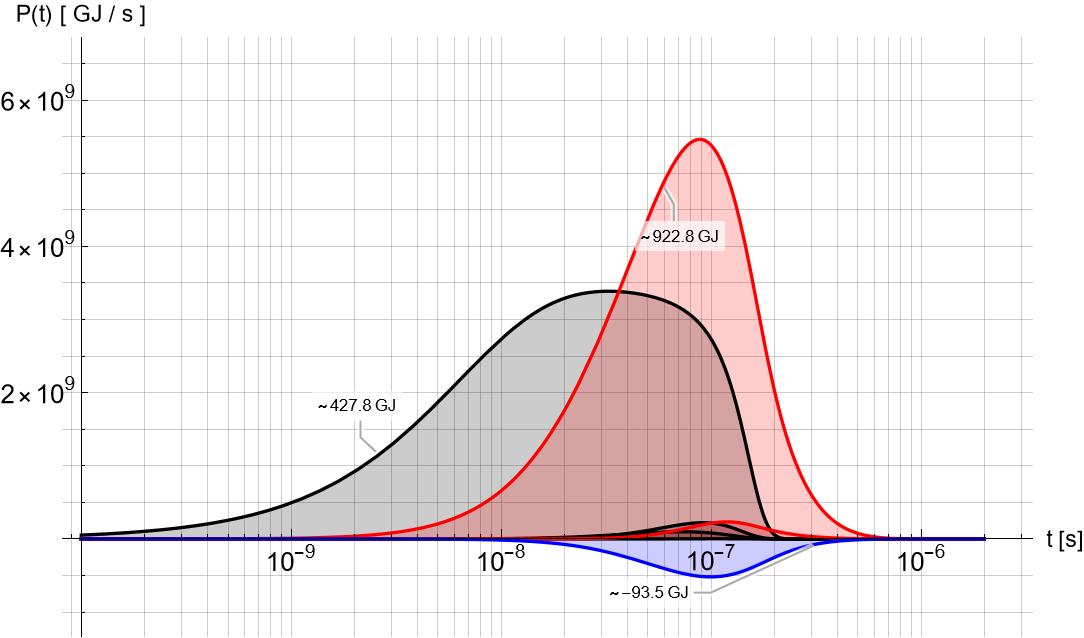}
	\end{center}
	\caption{Power curve for each reaction in the ideal case  for 1 mole of fuel material as a function of time in GJ/s for injected neutrons with 0.24 MeV kinetic energy. The integral for the three largest curves is indicated. The total integrated energy is $\sim$1323 GJ.}
	\label{specpow}
\end{figure}

The specific power rate (= power/volume/time) for each reaction is defined as:
\be 
P(t) = \rho N_A Y_i(t) Y_j(t) [(i,j),k]Q \;.
\ee
If we convert the Q value from MeV such that $P$ is in GJ/(s m$^3$), we can get the power as a function of time upon multiplying for the molar volume as plotted in Fig. \ref{specpow} for the present case.  The overall molar energy produced, obtained by time integration, is reported on the figure. Positive and negative curves represent produced and absorbed energy, respectively. If the total balance is positive, as in this case, we actually have an energy gain, indicated in the picture.
Off this purely theoretical figures, one should subtract, at least, the energy needed to accelerate neutrons up to the given kinetic energy and then multiply by some efficiency factor. Even at 0.1\% level, this energy gain is impressive, but one must bear in mind that this is the ideal case, in which no energy losses are included. Very different will be the outcome of the realistic case in the next section.

Now, with the aim of finding the optimal conditions to run this reactions, we have repeated the calculations several times for several different initial energies and neutron injection rates.
	\label{totspec}

At lower neutron energies, the cross-section for the triggering reaction (see insert in Fig. \ref{cs}) is larger without too much energy being ``stored" in the accelerated beam. The amount of liberated energy is always the same, just the reactions proceed more slowly (tenths of a ms instead of $\mu$s), but the burning is more efficient, i.e. the conversion into $\alpha$ particles reaches higher percentages. This looks like an ideal setting for the nuclear fusion cycles.
The rate of the first reaction is essentially flat at low neutron energies, therefore, despite we started our calculations by setting the neutron energy to the maximum, any lower energy would do with the same efficiency, only slower in time, therefore the liberated power would be less.

\section{Results for the realistic case}\label{real}
Fast charged particles generated by the first and subsequent reactions must travel through solid matter (before it eventually blows up), therefore they will loose energy by ionizing the surrounding material. This phenomenon can be modeled with the well-known Bethe-Bloch formula for the so-called stopping power as it appears on books of Krane \cite{Krane} and  Segr\'e \cite{Segre}. Very useful is the review of Seltzer \& Bergen \cite{SelBe}, in the present case the high Z correction of Barkas and Bloch is not implemented, therefore the approach is not valid below a certain minimum energy/velocity. At the energies that we are interested in, these charged particles do not typically travel more than a few tens of $\mu$m, exhausting most of their kinetic energy before the cycles of reactions can propagate beyond the first step. This is a mishappening for our purpose, but one might get enough energy already from the very first step.
Therefore, the set of equations of Appendix A must be supplemented by a new set of 8 non-linear equations (see appendix B) and the kinetic energies of all particles involved must be promoted to new variables (as they are changing along with the reaction dynamics). The solution of the system is now harder to get and a set of additional boundary conditions must be supplemented, namely the initial kinetic energy, as determined by Q-value considerations, and the request that kinetic energy does not go negative: this is not an issue as long as one insures that the solutions behave smoothly such that when the kinetic energy approaches zero (and we enter in the regime where the Bethe-Bloch loses applicability) we can forcibly set it to zero.

The neutron deceleration (due to impacts with the dense medium) and diffusion is not yet implemented, because is thought to be of minor relevance to the dynamics of the reaction network.

The solution of the system of equations works reasonably well in a certain range of neutron intensities, that luckily comprehend a region where the reaction happens to generate an interesting energy output.
The typical nuclear burning time for a mole of crystal is about or just less than 1 ns.

\begin{figure}[t!]
	\begin{center}
		\includegraphics[clip=,width=.7\textwidth]{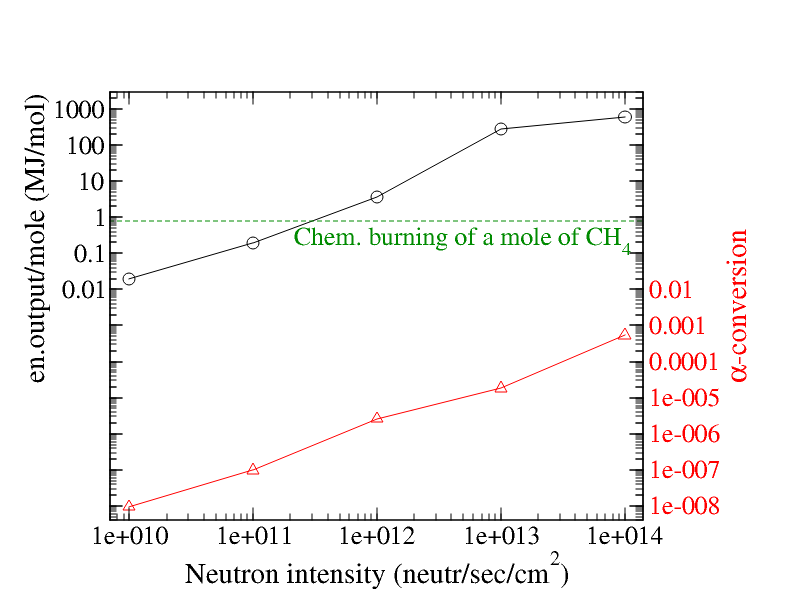}
	\end{center}
	\caption{Left, black scale: energy output for a mole of crystal burned, to be compared with the energy obtained from (chemically) burning a mole of methane. Right, red scale: the alpha conversion fraction.}
	\label{powstop}
\end{figure}

The figure \ref{powstop} shows two scales: on the left, in black the energy output (in MJ/mol) indicates that these reactions should still be advantageous if compared with chemical burning of a mole of CH$_4$ fuel, by a factor 1000 at the neutron beam intensity of $10^{14}$ neutrons/cm$^2$/sec. On the right, in red, the alpha conversion factor is the fraction of nuclear fuel converted into alpha particles. This parameter can be taken as an indicator of burning efficiency. Clearly, the neutron beam should be intense and well concentrated to generate usable thermal outputs from the mass of nuclear fuel. One might wonder about the instantaneous ablation of the heated fuel. We guess that this could be tampered by letting the neutron hit a much larger target than just a mole of material. Of course one should experimentally investigate the reactions by a slowly increasing the beam intensity.

\section{Conclusions and a call to action}
Our calculations demonstrate that the Jetter's and Post's nuclear fusion cycles involving $^6$Li deserve further attention and especially they should undergo a serious program of experimental testing, in which, essentially a neutron beam with variable energy (but not necessarily high) should be directed to a tiny bit of fuel material, surrounded by other fuel and a calorimeter to assess the effective yields  \footnote{Very recently, we came to know the opinion of experts in the field of neutron measurements, that a calorimeter is very delicate device to build. Probably some activation or other indirect measurements could be more favorable setups.}.
Clearly, there is a fraction of elastically scattered neutrons that do not interact with the fuel, one should experiment with various geometries to retain as many neutron within the bulk of fuel as possible.
Calculations including the Bethe-Bloch stopping power of charged particles produced in the primary reaction into the crystal itself show that it is still worth burning this fuel with neutrons, even after taking into account energy losses that significantly alter the ideal reaction dynamics. Essentially, we find that the secondary and ancillary reactions do not significantly contribute to the total energy release, unless a very intense neutron beam is used. 
One should also note that the realistic case depicted in Fig. \ref{powstop} is definitely not ideal and that the flux on the rightmost part of the figure is a huge one, therefore some practical limitations might be in order.  

In parallel, additions to the theoretical calculations (side reactions, other forms of energy loss, etc.) might reveal new aspects and they should be implemented and further investigated. Other reactions should be added to the list, in some cases, especially when radioactive and rare tritium is involved either as a target or as a projectile, data compilations are scarce and a call to action to low-energy experimental nuclear physicists should be encouraged.

\appendix

\section{Evolution equations and technicalities}

The set of ordinary differential equations that describes the abundance evolution of the system reads
\begin{align}
{\dot{Y}_{n_1}} &= i - Y_{n_1} Y_{\mathrm{Li}} \left[ \mathrm{Li}(n_1,\alpha) \right] -\lambda_{n_1}Y_{n_1}
\\
{\dot{Y}_{n_2}} &= Y_{\mathrm{D}} Y_{\mathrm{T}} \left[ \mathrm{D}(\mathrm{T},n_2) \right]- Y_{n_2} Y_{\mathrm{Li}} \left[ \mathrm{Li}(n_2,\alpha) \right] \nonumber
\\&	- Y_{n_2} Y_{\mathrm{D}} \left[ n_2(\mathrm{D},2n) \right] -\lambda_{n_2}Y_{n_2}
\\
{\dot{Y}_{n_3}} &= Y_{\mathrm{D}} Y_{n_1} \left[ n_1(\mathrm{D},2n) \right] -\lambda_{n_3}Y_{n_3}
\\
{\dot{Y}_{n_4}} &= Y_{\mathrm{D}} Y_{n_2} \left[ n_2(\mathrm{D},2n) \right] -\lambda_{n_4}Y_{n_4}
\\
{\dot{Y}_{\alpha_1}} &= Y_{n_1} Y_{\mathrm{Li}} \left[ \mathrm{Li}(n_1,\alpha_1) \right]+ Y_{n_2} Y_{\mathrm{Li}} \left[ \mathrm{Li}(n_2,\alpha_1) \right]
\\
{\dot{Y}_{\alpha_2}} &= Y_{\mathrm{D}} Y_{\mathrm{T}} \left[ \mathrm{D}(\mathrm{T},\alpha_2) \right] 
\\
{\dot{Y}_{\alpha_3}} &= Y_{p_1} Y_{\mathrm{Li}} \left[ \mathrm{Li}(p_1, \mathrm{He} ) \right] + Y_{p_2} Y_{\mathrm{Li}} \left[ \mathrm{Li}(p_2, \mathrm{He} ) \right] 
\\
{\dot{Y}_{\alpha_4}} &= Y_{\mathrm{D}} Y_{\mathrm{He}} \left[ \mathrm{He}(\mathrm{D},p) \right] 
\\
{\dot{Y}_{\mathrm{D}}} &= - Y_{\mathrm{D}} Y_{\mathrm{T}} \left[ \mathrm{D}(\mathrm{T},n_2) \right] - Y_{\mathrm{D}} Y_{\mathrm{He}} \left[ \mathrm{He}(\mathrm{D},p) \right]
	- Y_{\mathrm{D}} Y_{n_2} \left[ n_2(\mathrm{D},2n) \right]
\\
{\dot{Y}_{\mathrm{T}}} &= Y_{n_1} Y_{\mathrm{Li}} \left[ \mathrm{Li}(n_1,\alpha) \right]+ Y_{n_2} Y_{\mathrm{Li}} \left[ \mathrm{Li}(n_2,\alpha) \right] \nonumber
\\&	- Y_{\mathrm{D}} Y_{\mathrm{T}} \left[ \mathrm{D}(\mathrm{T},n_2) \right]-\lambda_{T}Y_{T}
\\
{\dot{Y}_{\mathrm{He}}} &= Y_{p_1} Y_{\mathrm{Li}} \left[ \mathrm{Li}(p_1, \mathrm{He} ) \right] + Y_{p_2} Y_{\mathrm{Li}} \left[ \mathrm{Li}(p_2, \mathrm{He} ) \right] 
	- Y_{\mathrm{He}} Y_{\mathrm{D}} \left[ \mathrm{He}(\mathrm{D},p_2) \right]
\\	
{\dot{Y}_{\mathrm{He0}}} &= \lambda_{T}Y_{T}
\\
{\dot{Y}_{p_1}} &= Y_{\mathrm{D}} Y_{n_2} \left[ n_2(\mathrm{D},2n) \right] - Y_{p_1} Y_{\mathrm{Li}} \left[ \mathrm{Li}(p_1, \mathrm{He} ) \right] 
\\
{\dot{Y}_{p_2}} &= Y_{\mathrm{D}} Y_{\mathrm{He}} \left[ \mathrm{He}(\mathrm{D},p_2) \right]- Y_{p_2} Y_{\mathrm{Li}} \left[ \mathrm{Li}(p_2, \mathrm{He} ) \right] 
\\
{\dot{Y}_{p_0}} &= \lambda_{n_1}Y_{n_1} +\lambda_{n_2}Y_{n_2}+\lambda_{n_3}Y_{n_3}+\lambda_{n_4}Y_{n_4}
\\
{\dot{Y}_{\mathrm{Li}}} &= - Y_{\mathrm{Li}} Y_{n_1} \left[ \mathrm{Li}(n_1,\alpha_1) \right]- Y_{\mathrm{Li}} Y_{n_2} \left[ \mathrm{Li}(n_2,\alpha_1) \right] \nonumber\\ &- Y_{p_1} Y_{\mathrm{Li}} \left[ \mathrm{Li}(p_1, \mathrm{He} ) \right] - Y_{p_2} Y_{\mathrm{Li}} \left[ \mathrm{Li}(p_2, \mathrm{He} ) \right]
\end{align}
where the all the $Y_i$ (in mol/g) depend on time, the dot indicates time-derivative and $\left[ j(k,i) \right] = \rho N_A \ev{\sigma v}_{jk,i}$ has been used as a shorthand notation for the rate (in g/(mol$\cdot$s)). 
The neutron injection rate, $i$ (in mol/g$\cdot$s), is proportional to the neutron flux (or to  neutron current). The scattering of neutrons is not included.
Albeit negligible, the $\beta$ decay of tritium $^3$H $\rightarrow$ $^3$He +e$^-+\overline{\nu}_e $ and neutrons n $\rightarrow$ p +e$^-+\overline{\nu}_e$(considered free) is included. The released energy, $Q=18.6$ keV and $Q=0.782$ MeV respectively, is carried away by leptons and eventually thermalizes, leaving behind an essentially motionless daughter nucleus ($^3$He and $p$) that does not further propagate the reactions. These leftover ashes are collected in sink terms, $Y_{p_0}$ and $Y_{He0}$ and the typical outcome of the calculations indicates that these are negligible.
For the coupling reaction, (\ref{knock}), experimental data points terminate at about 12 MeV, while we need the rate at 14.6622 MeV. In absence of other strategies, this value has been extrapolated to $\sim 4.605\cdot 10^{-7}$ $m^3/s$ $mol$.
The initial conditions on molar fractions are taken as $Y_{\mathrm{D}} = Y_{\mathrm{Li}} = 0.125 $(mol/g) to comply with the condition $\sum A_i Y_i=1$.
This system of differential equations does not present any stiffness or singularity, and it is always smoothly solved as long as one keeps away from the limits of definition of the cross-section data.

\section{Stopping power equations}
Each charged species loses energy, the loss being faster when the particle slows down, therefore the 8 equations for charged particles produced by the reaction cycles  $n_i=\{ T, \alpha_1 , \alpha_2, \alpha_3, \alpha_3 , p_1, p_2, He\}$, are written in terms of the kinetic energy $K$ as 
\be
{\dot{K}_{n_i}} = -\frac{dE}{dx}(K_{n_i}(t))
\ee
where the energy loss is given in terms of velocity $v(K)$ and $\beta=v/c$, by
\be
-\frac{dE}{dx}=\Bigl(\frac{ze^2}{4 \pi \varepsilon_0}\Bigr)^2\frac{ 4\pi  Z \rho  N_A}{ A m_e v^2} \Bigl[ ln \Bigl( \frac{ m_e v^2}{I}\Bigr) - ln(1-\beta^2) - \beta^2 \Bigr] \label{BB}
\ee
Please refer to standard textbooks for nomenclature \cite{Krane, Segre}. 
Neutrons are also subject to energy loss by impact\cite{Segre}, but this is not yet implemented as it is supposed to be a minor effect.

\section*{Author contributions and Acknowledgments}
L.F. (60\%) found the case in literature, envisioned the problem, wrote the first draft of the code, checked each version and wrote the manuscript. A.F.L.L. (30\%) searched and collected data from databases, worked to the code for his M.Sc. thesis \cite{Lope} and prepared some figures, G. A. (10\%) contributed to the initial stage and prepared some data sets for his B.Sc. thesis.
E.F. ran GEANT simulations to confirm ranges and rates.

We thank K.Starosta and H.Asch (Simon Fraser Univ., B.C., Canada) for useful discussions. 

\section*{Data availability statement}
Data and code \cite{Code} will be made available through the open-access {\it Research Data Unipd} archive of the University of Padova \cite{RDunipd}.

\end{document}